# Methane Incorporation into Liquid Fuel by Non-Equilibrium Plasma Discharges

Chong Liu[1, 2], Ivan Chernets[1, 3], Hai-Feng Ji[4], Joshua Smith[4], Alexander Rabinovich[1], Danil Dobrynin[1*], Alexander Fridman[1, 3]

1. Drexel Plasma Institute 200 Federal St., Suite 500 Camden, NJ 08103, USA
2. Department of Electrical and Computer Engineering, Drexel University, 3141 Chestnut ST, Philadelphia, PA 19104, USA
3. Department of Mechanical Engineering and Mechanics, Drexel University, 3141 Chestnut ST, Philadelphia, PA 19104, USA
4. Department of Chemistry, Drexel University, 3141 Chestnut St, Philadelphia, PA 19104, USA

**Abstract:** The conventional ways of processing natural gas into more efficient and economical fuels usually either have low conversion rate or low energy efficiency. In this work, a new approach of methane liquefaction is proposed. Instead of direct treatment of only natural gas, plasma activated methane is reacting with liquid fuel. In this way, methane molecules are directly incorporated onto liquid fuel to achieve liquefaction. Nanosecond-pulsed dielectric barrier discharge and atmospheric pressure glow discharge are used here to ensure no local heating in gas bubbles. Effects of both discharges on methane reaction with liquid fuel are investigated, mass and chemical changes in liquid are observed. Preliminary results show fixation of methane in liquid fuel.

## 1. Introduction

With the increasing discovery of shale gas resources all over the world, cheap natural gas has become a valuable alternative fuel. The usage of natural gas in power generation, heat, transportation, fertilization and other applications is increasing. The main obstacle that makes it a less desirable option for industry is its gaseous state. Intensive research efforts in the area of direct methane conversion into more valuable liquid hydrocarbons have been made. More recently, plasma chemical conversion of natural gas has been investigated, including generation of syngas and gaseous hydrocarbons [1-11].For example, Wang and Xu [1] have shown that in presence of catalysts, it is possible to achieve more than 60% methane conversion into gaseous C2 hydrocarbons using cold plasma reaction. They argue that the major process is formation of CH3 radicals with generation of ethane and ethylene. Sentek et al [2] have tested a number of catalysts in combination with DBD plasma for conversion of methane – CO2 mixtures into higher hydrocarbons and alcohols. Similar studies by Liu et al have been conducted [3]; they demonstrated possibility of generation of liquid fuel using DBD discharge ignited in methane-CO2 mixture. In contrast, Indarto and co-authors could not detect any methane liquefaction effect of gliding arc discharge [4], with hydrogen and acetylene being the main products.

---

[*] Corresponding author: dvd34@drexel.edu

Here we propose another possible mechanism of methane liquefaction by its incorporation into liquid hydrocarbons stimulated by non-thermal plasma. In general, incorporation of methane molecules may follow two possible reaction paths: saturation of a carbon double bond, or polymerization. The reaction that occurs during this process – saturation of aromatic hydrocarbons $R_1 = R_2 + CH4 \rightarrow HR_1 - R_2CH_3$ ($\Delta H = -0.5 \, eV/mol$) is exothermic and energy cost is not more than 0.3 eV/mol of CH4. All other reactions between hydrocarbons and non-thermal plasma (such as polymerization and dissociation) are strongly endothermic and therefore not interesting for effective direct liquefaction process. By applying non-equilibrium discharge the local heating can be avoided and saturation will be the prevailing chemical process, thus achieving better energy efficiency.

In this work this new method of methane liquefaction is implemented and tested. Mass balance and chemical analysis of both gaseous methane/nitrogen mixture and liquid samples were performed to quantify the incorporation process.

## 2. Materials and experimental methods

In this study we have used two plasma systems for incorporation of methane/nitrogen mixtures into liquid hydrocarbons: atmospheric pressure nanosecond-pulsed DBD and atmospheric pressure glow discharge. Both discharges were ignited inside gaseous bubbles fed through a liquid hydrocarbons in order to increase reaction area as well as mixing.

### 2.1. Atmospheric pressure nanosecond-pulsed DBD system

Dielectric barrier discharge was ignited in a coaxial cylinder reactor (Figure 1). For that, a quartz cylinder (3.7 cm ID, 4 cm OD, and 15.2 cm long) was filled with liquid hydrocarbons, while gas mixture fed through a dielectric UHMW polyethylene tube (3.5 cm OD) dispersed into ~1mm diameter bubbles using ceramic airstone fixed at the bottom of the system. Copper woven wire mesh (0.4 mm wire diameter) was wrapped around inner and outer surfaces of the quartz tube served as two electrodes. The inner mesh electrode was connected to the high voltage output of the power supply, and the outside one was grounded. High voltage (+10.1kV pulse) pulses with duration of 10 ns (90% amplitude), rise time of 2 ns and frequency of 490 Hz were applied using FIDTech power supply via a 15 m long coaxial cable RG393/U. Calibrated back current shunt was mounted in a middle of the cable and was used for control of applied voltage and power measurements (single pulse discharge energy was measured to be 3 mJ when power supply is set to +10.1kv). Methane and nitrogen gases were fed into the system using Alicat M-series mass flow controller at the rate of 0.2 slpm.

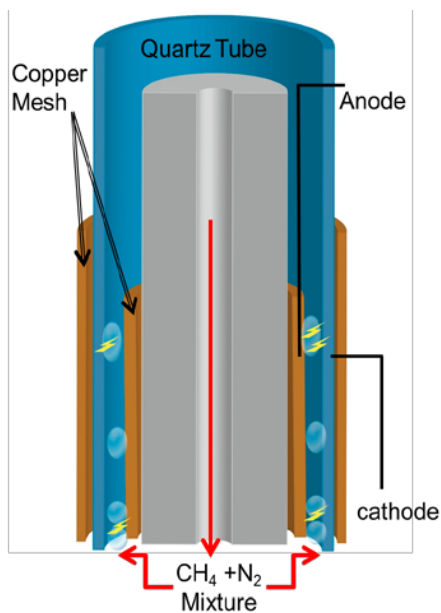

Figure 1. Dielectric barrier discharge reactor system.

## 2.2. Atmospheric pressure glow (APG) discharge system

Schematic of APG plasma process is shown in Figure 2. Discharge was initiated between ground rod electrode and tubular high voltage electrode through which gas mixture was dispersed as bubbles into liquid hydrocarbon mixture. The APG reactor consisted of a PVC tube with OD 42 mm, ID 31 mm and 4 electrode pair which were arranged in two levels 15mm apart. Each electrode pair had a ground and a high voltage (HV) electrodes. The ground electrode consisted of a copper rod with 3.2 mm diameter and the HV electrode was made off a copper tube with OD=3.2 mm and ID=0.9 mm. Each HV electrode was connected to a HV power supply (PS, Universal Voltronics BRS 10000) through 10 MOhm resistor. The gas mixture was supplied through a HV electrode consisted of 0.27 slpm of $N_2$ and 2.7 slpm of $CH_4$ and flow was controlled by Alicat mass flow controllers MC-5SLPM-D. The liquid fuel was low sulfur diesel (weight 50 gram). The supplied gas created a bubble between two electrodes where discharge occurred. Voltage and current were measured by DPO3014 oscilloscope and Pearson 6165 current monitor.

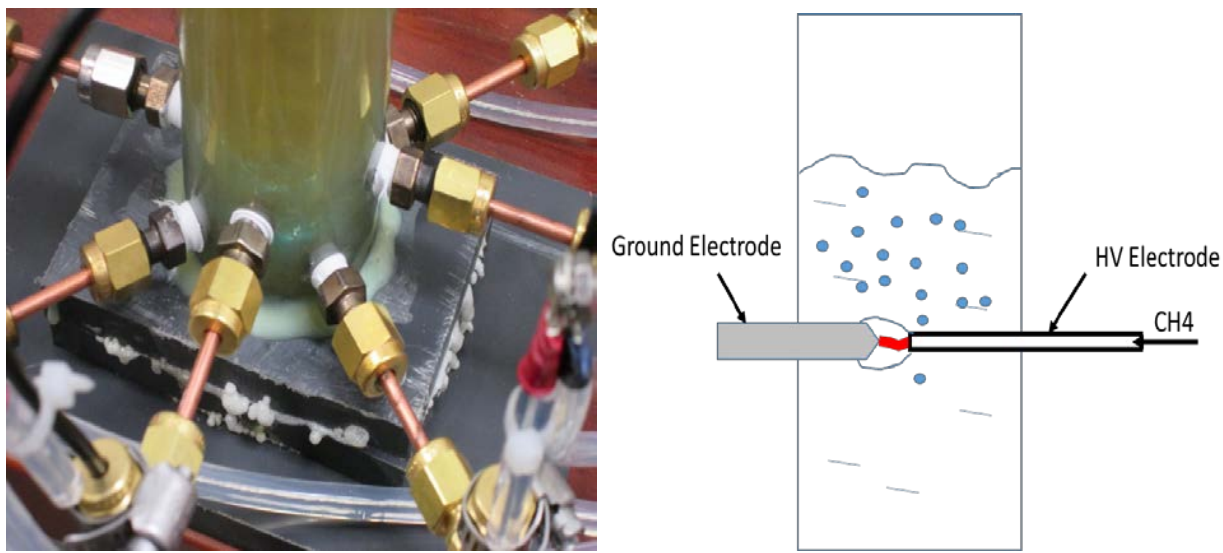

*Figure 2. Atmospheric Pressure Glow Discharge plasma system*

## 2.3. Chemicals

In our experiments, we have used low sulfur diesel; and 1-methylnaphthalene (M56808 ALDRICH 95%) was used as replacement of diesel. The molecular structure of 1-methylnaphthalene is shown in Figure 3. Similar to diesel, 1-methylnaphthalene has a number of C-H double bonds, as well as two hydrocarbon rings.

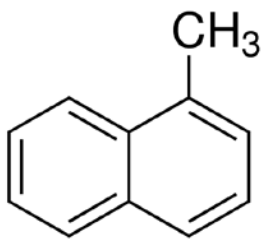

*Figure 3. Molecular structure of 1-methylnaphthalene*

### 2.4 Gas and liquid phase analysis

Here we have used several techniques for analysis of chemical processes occurring during methane incorporation into liquid hydrocarbons. For analysis of gas chemistry, we have utilized gas chromatography technique (490 Micro chromatograph, Agilent Technologies); liquid phase analysis was performed using IR absorption (NICOLET 8700 with smart-iTR sampling accessory for liquid sample): all spectrum are acquired in resolution of 4 nm and averaged over 32 repeated measurements. Nuclear magnetic resonance spectroscopy (NMR) was also performed.

## 3. Experimental results and analysis

### 3.1. Mass balance

One of the simplest ways to demonstrate methane incorporation into liquid hydrocarbons is mass balance analysis. For that, the weight of the DBD reactor containing initially 30 ml of liquid diesel was measured before and after experiments. During experiment, methane/nitrogen mixture at rate of 0.2 slpm (0.1 slpm of methane, and 0.1 slpm of nitrogen) was fed through the DBD discharge. Weight was measured every 10 minutes using a Satorius Entris balance, which has resolution of 20 mg; 100 records was taken and averaged in each measurement to minimize the instrumental error. Another treatment of the same amount of fresh diesel sample with 0.2 slpm nitrogen flow was done as a control experiment. The results are shown in Figure 4.

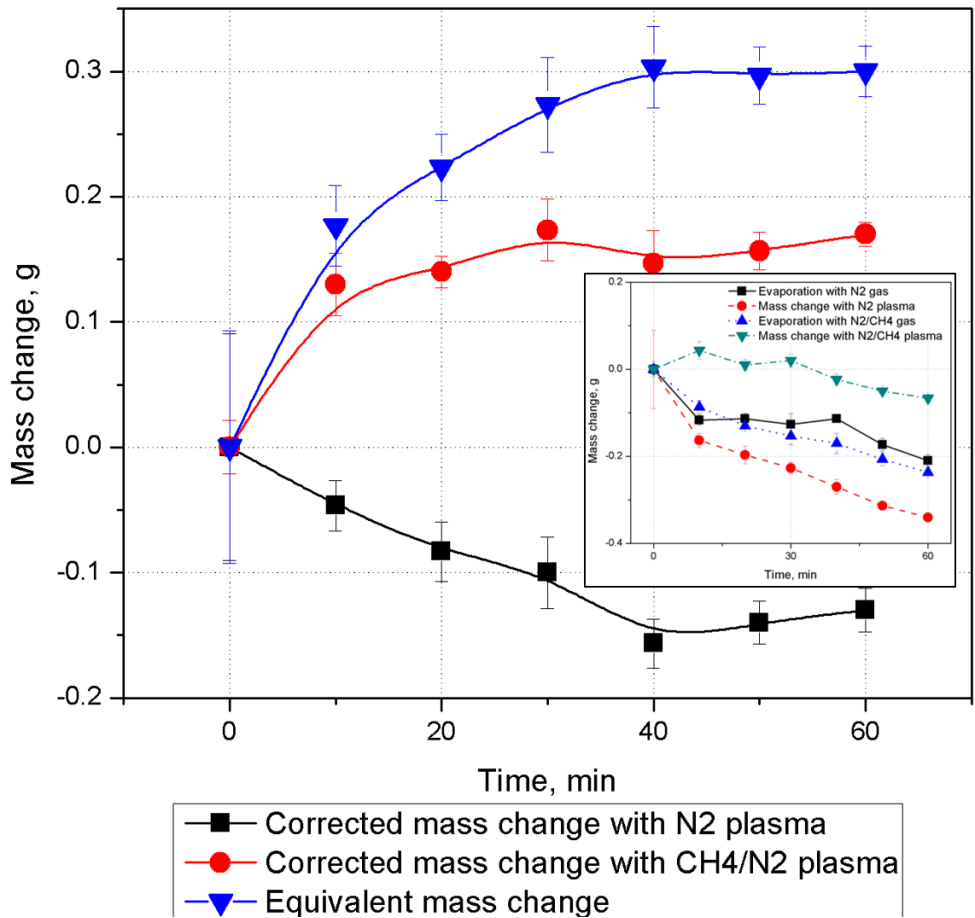

*Figure 4. Weight loss over time with respect to different gases corrected for evaporation losses due to bubbling (insert: raw data showing uncorrected mass change due to effects of plasma and evaporation losses)*

Here we have observed absolute mass increase indicating possible methane incorporation effect. The mass change is calculated as mass increase corrected for evaporation losses as shown in Figure 4: after an hour of DBD treatment with methane-nitrogen mixture a 0.17 g of mass increase was observed. Taking plasma power (1.5 W) into consideration, the energy cost of incorporation os 0.17 g of methane in 60 minutes of treatment may be calculated as 5.1 ev/molecule, with conversion rate of ~4%. At the same time, assuming that the liquid removal is about the same for plasma ignited nitrogen and nitrogen-methane mixture, and that all removed liquid may be collected downflow with appropriate condensation unit, we can calculate "equivalent mass increase" as shown in Figure 4. Corresponding energy cost of incorporation of 0.3 g of $CH_4$ in 60 min is 2.9 eV/molecule with 7% conversion rate. Interestingly, we notice saturation of mass increase nature of which is still unknown, and may be related to overall temperature increase, or chemical changes in liquid phase. Calculations of energy cost and conversion rates for the first 10 min of treatment (before saturation) are the following: 1.1 eV/molecule at 18% for absolute mass change corrected for evaporation, and 0.8 eV/molecule at 23% for "equivalent" mass increase.

### 3.2. Gas phase analysis

To further understand the effect of methane incorporation into liquid fuel, gas chromatography (GC) was performed to analyze the gas composition changes after plasma treatment in both setups.

In the case of DBD plasma system, a premixed methane/nitrogen mixture (50% $CH_4$ and 50% $N_2$) from Airgas was fed through 1-methylnaphthalene. Gas flow rate was set to 0.2 slpm, and gas samples were collected after the reactor for analysis using GC (490 Micro, Agilent Technologies). The results are presented in Table 1.

*Table 1. Gas composition before and after DBD treatment of CH4/N2 mixture in 1-methylnaphthalene*

|  | $N_2$, vol. % (L) | $CH_4$, vol. % (L) | $C_2H_6$, vol. % (L) | $C_2H_2$, vol. % (L) | $C_2H_4$, vol. % (L) | $H_2$, vol. % (L) |
|---|---|---|---|---|---|---|
| **Without Discharge** | 52.99% (0.106) | 46.99% (0.094) | <0.02% ($3\times10^{-5}$) | Not detectable | Not detectable | Not detectable |
| **60min of treatment** | 53.1% (0.106) | 46.59% (0.093) | <0.09% ($17\times10^{-5}$) | <0.02% ($3\times10^{-5}$) | Not detectable | 0.2% ($4\times10^{-4}$) |

Experiments with APG treatment of low sulfur diesel were performed using gas mixture of $CH_4$ (2.7L/min) and $N_2$ (0.27 L/min). Plasma parameters were: voltage – 2.4 kV, current – 0.62 mA. Exhaust gas samples taken every minute during 5 min experiment time were analyzed by GC (490 Micro, Agilent Technologies). Experimental results of gas chromatographic analysis are shown in Table 2.

*Table 2. Gas composition before and after APG treatment of CH4/N2 mixture in diesel*

| Treatment time, min | $N_2$, vol. % (L) | $CH_4$, vol. % (L) | $C_2H_6$, vol. % (L) | $C_2H_2$, vol. % (L) | $C_2H_4$, vol. % (L) | $H_2$, vol. % (L) |
|---|---|---|---|---|---|---|
| 0 | 8.481 (0.27) | 91.513 (2.7) | $6\times10^{-3}$ ($19\times10^{-5}$) | 0 | 0 | 0 |
| 1 | 10.775 (0.27) | 89.201 (2.235) | $7\times10^{-3}$ ($18\times10^{-5}$) | $2\times10^{-3}$ ($5\times10^{-5}$) | 0 | $15\times10^{-3}$ ($38\times10^{-5}$) |
| 2 | 10.710 (0.27) | 89.263 (2.250) | $7\times10^{-3}$ ($18\times10^{-5}$) | $3\times10^{-3}$ ($8\times10^{-5}$) | 0 | $17\times10^{-3}$ ($43\times10^{-5}$) |
| 3 | 10.563 (0.27) | 89.409 (2.285) | $7\times10^{-3}$ ($18\times10^{-5}$) | $3\times10^{-3}$ ($8\times10^{-5}$) | 0 | $18\times10^{-3}$ ($46\times10^{-5}$) |
| 4 | 10.613 (0.27) | 89.387 (2.274) | $7\times10^{-3}$ ($18\times10^{-5}$) | $3\times10^{-3}$ ($8\times10^{-5}$) | 0 | $20\times10^{-3}$ ($51\times10^{-5}$) |
| 5 | 10.450 (0.27) | 89.519 (2.313) | $8\times10^{-3}$ ($21\times10^{-5}$) | $3\times10^{-3}$ ($8\times10^{-5}$) | 0 | $21\times10^{-3}$ ($54\times10^{-5}$) |

Significant decrease in CH4 volume (~0.4L for 5 min of treatment) could not be explained by methane dissociation with production of H2, C2H2 and C2H6. The only reasonable explanation of this decrease would be CH4 incorporation into liquid diesel. Energy cost calculated from these parameters is quite low - 0.32 kW-h/m3 of CH4 that confirm the prevailing role of exothermic process of aromatic rings saturation.

### 3.3. FTIR analysis of liquid fuel

DBD plasma experiments were performed at 0.1 slpm flow rates of both methane and nitrogen. Mixture of 1-methylnaphtalene and hexadecane was used. The mixture was treated with methane/nitrogen mixture in the same DBD reactor for an hour. Inferred absorption spectrums are recorded and compared (with OMNIC software) for each sample using NICOLET 8700 (with smart-iTR sampling Accessory for liquid sample).

Spectra of the liquid sample before and after experiment are shown in Figure 5; there are significant differences around 2850-3100 cm$^{-1}$ (C-H bond absorption [12]) and 700-800 cm$^{-1}$ (phenyl ring function group [12]). The third graph in the figure shows the difference, indicating decrease in phenyl ring function group and unsaturated C-H bond with corresponding increase of saturated C-H bond. This suggests that discharge opens phenyl ring, and activated methane saturates carbon double bond. The same experiment is done with only DBD in nitrogen, even though there is decrease in phenyl ring group concentration but there is no obvious change in C-H absorption band.

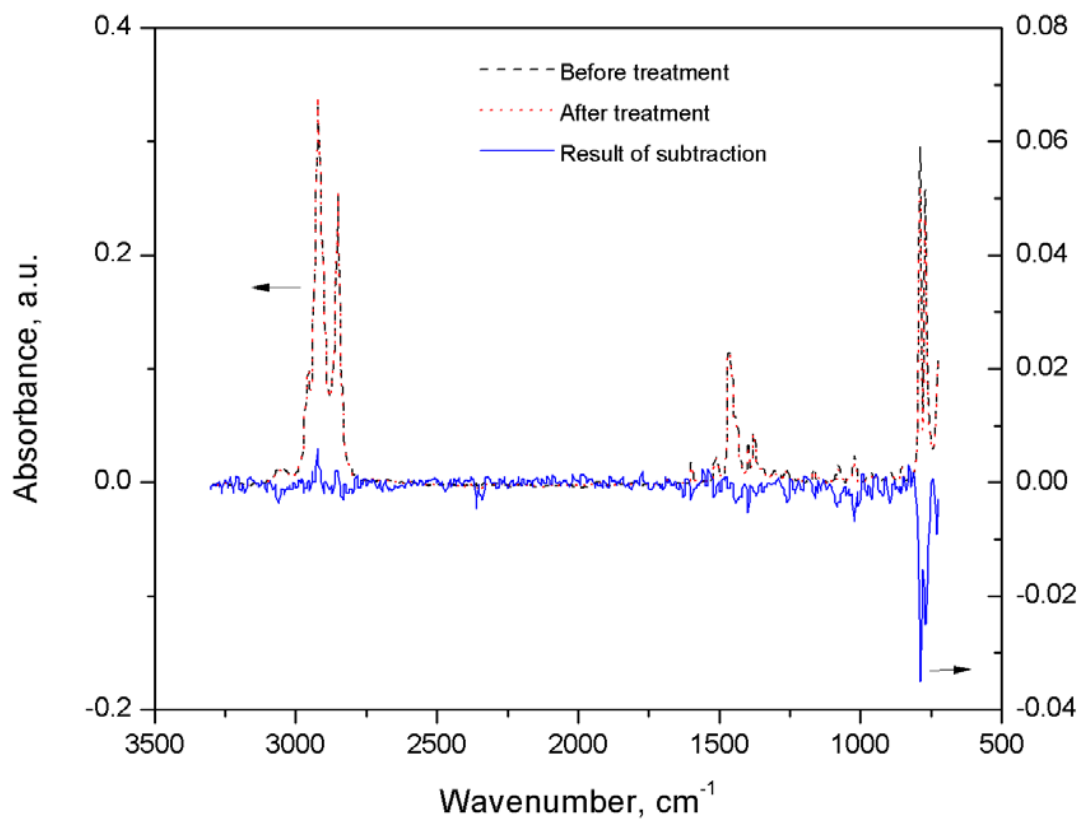

*Figure 5. Comparison of FTIR spectrum of liquid sample before and after treatment*

To quantify this chemical concentration change, absorption of saturated hydrocarbon bond was calibrated with respect to the concentration of 1-methylnaphthalene. Five sets of standard solution of 1-methylnaphthalene (97+%, ACROS Organics) and hexadecane with 1-methylnaphthanlene mass concentration from 100% to 92% were used, and the absorption intensity of 2750 ~ 2950cm$^{-1}$ was recorded. The characteristic peak intensity is proportional to the concentration of saturated C-H bond, and it also appears to be inversely proportional to the 1-methylnaphtanlene concentration. The absorption intensity is plotted against the saturated C-H bond concentration (calculated based on 3 moles of saturated C-H bond per mole of 1-methylnaphthalene; 16 moles of saturated C-H bond per mole of hexadecane). Starting with 45g of pure 1-methylnaphthalene, after 2 hours of treatment the liquid sample is taken out the reactor. The inferred absorption of the sample at 2750 ~ 2950cm$^{-1}$ is measured and fitted with the calibration graph in Figure 6. With 45g of initial liquid in the system, after 2 hours there are approximately 0.2 mol of carbon bonds were saturated.

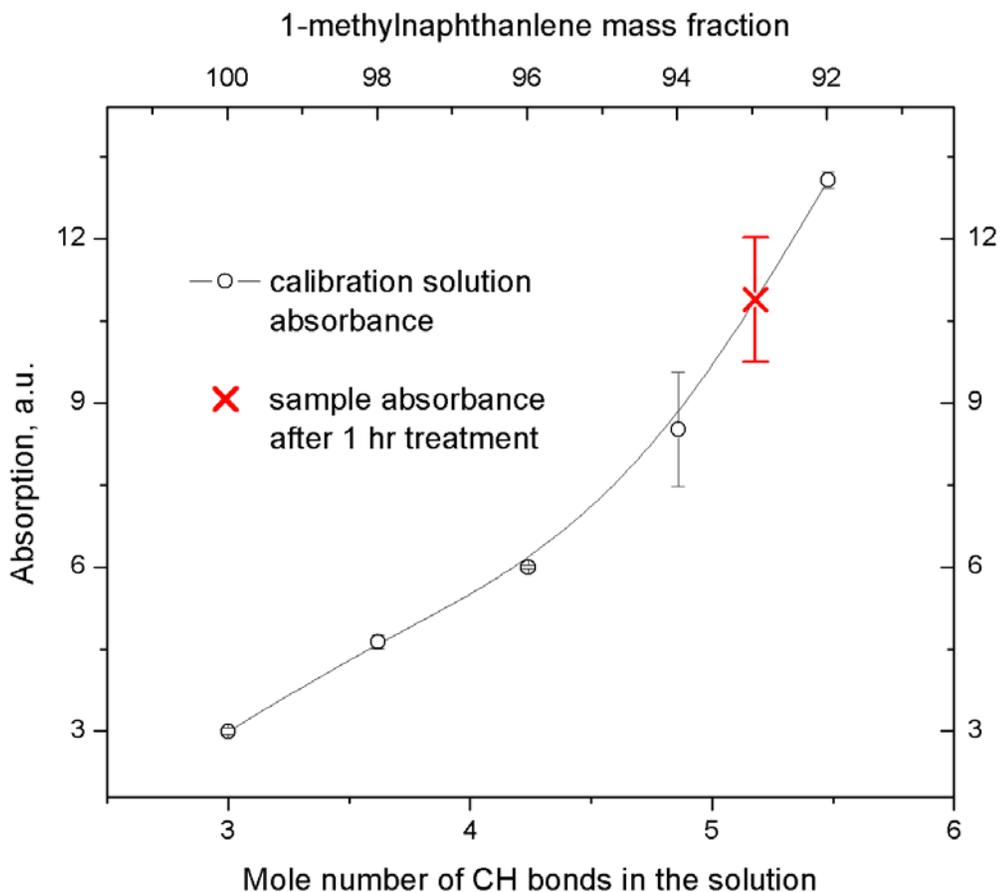

*Figure 6. Calibration of saturated C-H bond changes after treatment*

### 3.4. Nuclear Magnetic Resonance Spectroscopy (NMR) analysis of liquid hydrocarbons

Experimental tests were conducted for the plasma stimulated direct liquefaction of methane into a liquid aromatic and aliphatic compound mixture of methylnaphthalene and hexadecane, serving as a surrogate compound for liquid hydrocarbon fuel within the range of products found in diesel. The objectives of the tests were to determine (i) the selectivity of plasma activated natural gas for reactions with aromatics and aliphatics, (ii) the extent of aromatic ring saturation. The liquid products generated in the plasma system were analyzed through nuclear magnetic resonance spectroscopy (NMR), which preliminarily determined a significant selectivity to reactions with aromatics and ring saturation.

A mixture of 30% 1-methylnaphthalene and 70% hexadecane was treated with both discharges. This model solution is similar to diesel (contain phenyl ring in methylnaphthalene and saturated carbon chain in hexadecane). Results of NMR characterization (comparison of treated vs. untreated samples) are presented in Figure 7.

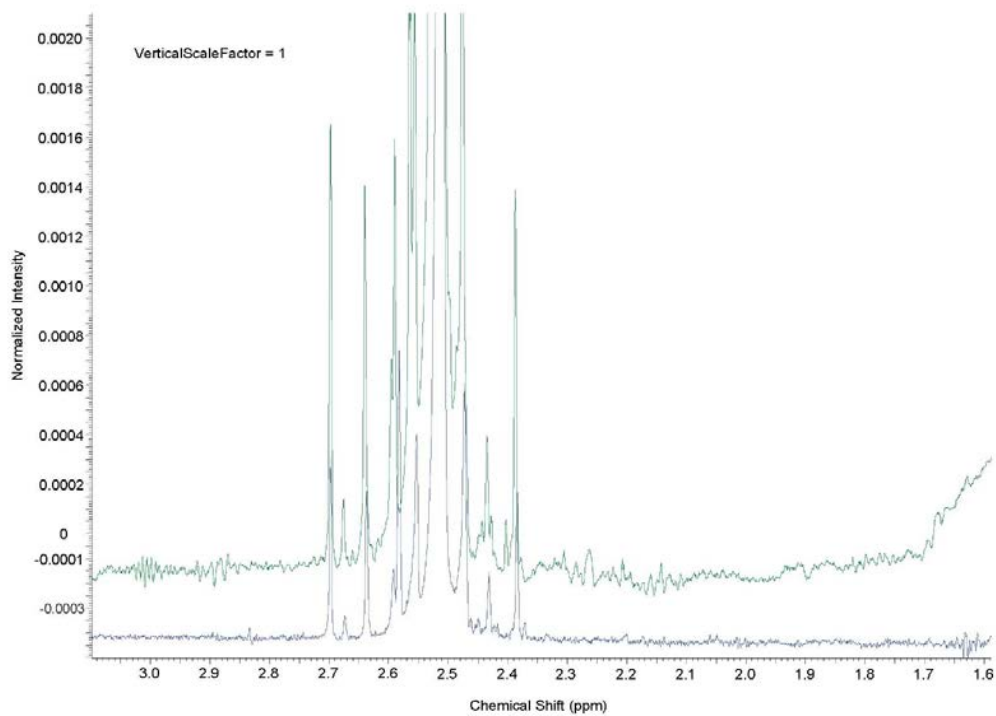
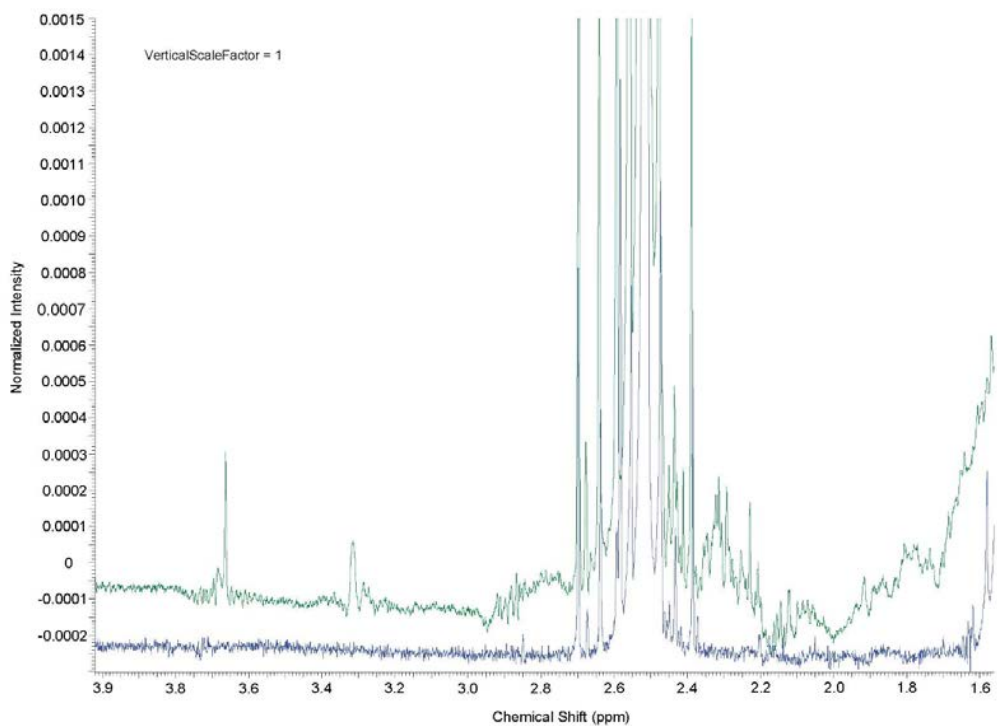

*Figure 7. NMR Spectrum for DBD (top) and APG (bottom) plasma treatment of liquid mixture (30% 1-methylnaphthalene C10H8, 70% hexadecane C16H34). Blue line – initial mixture; green line – after 1 hour treatment. Chemical shift in the range 1.6-2.4 shows phenyl-aliphatics.*

Preliminary analysis of the spectra shows that in both DBD and APG treatment portion of methane incorporated in liquid has been spent in saturation of aromatic rings, and much smaller portion has been used in polymerization of aliphatic compounds (hexadecane). About 85-90% of aromatic ring saturation corresponds to saturation of the first ring of 1-methylnaphthalene. About 10-15% of 1-methylnaphthalene has been converted into aliphatic compounds.

### 3.5. Analytical modeling of methane incorporation into liquid hydrocarbons using non-thermal plasma

For $CH_4$ (methane) incorporation into liquid fuel, we consider a high aromatic composition including:

- 70% aromatic molecules (e.g. anthracene $C_{14}H_{10}$)
- 30% saturated molecules (e.g. dodecane $C_{12}H_{26}$)

From a chemistry description, the process of attaching a molecule of methane to either of these types of molecules can be represented by the equations:

*Polymerization of saturated hydrocarbons:*

$$R - H + CH_4 \rightarrow R - CH_3 + H_2, \quad \Delta H = 0.6 \text{ eV/mol} \quad (1)$$

*Saturation of aromatic hydrocarbons:*

$$R_1 = R_2 + CH_3 \rightarrow HR_1 - R_2 CH_3, \quad \Delta H = -0.5 \text{ eV/mol} \quad (2)$$

Although the saturation reaction for aromatic molecules is exothermic, there is a finite activation barrier to the reaction of 0.2 eV/mol. Both reactions can be effectively stimulated by vibrationally excited $CH_4$ in plasma.

In liquid compositions where both reactions take place, we define the parameter $S$ called the selectivity which represents the favorability of the saturation reaction over polymerization. Thus, the minimum reaction energy $A_{min}$ for the overall process can be expressed as a function of $S$ according to:

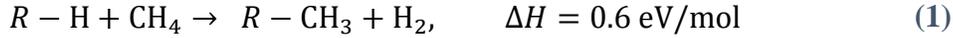
$$A_{min} = S\, 0.2 + (1 - S)0.6 \;\frac{\text{eV}}{\text{mol}} \quad (3)$$

This of course assumes 100% reaction efficiency, which is not the case in reality. Keeping in mind the effect of vibrational excitation in equations (1) and (2) and a realistic value of energy efficiency, the overall reaction efficiency $\Gamma$ can be parameterized as a function of $S$ according to the function:

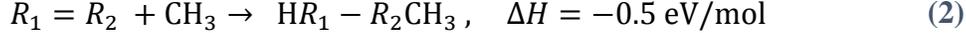
$$\Gamma = 0.2(1 + 3S) \quad (4)$$

The factor $(1 + 3S)$ is an empirically derived quantity to that theoretically minimum energy values are satisfied when $S = 0$ and $S = 1$. The function $\Gamma$ also accounts for the dependence of the conversion efficiency on the mechanism involved. Finally, the expression for the minimum energy required is:

$$A_{min} = \frac{0.6 - 0.4S}{\Gamma(S)} = \frac{0.6 - 0.4S}{0.2(1 + 3S)} \quad (5)$$

The graph of $A_{min}$ as a function of $S$ is shown in Figure 8.

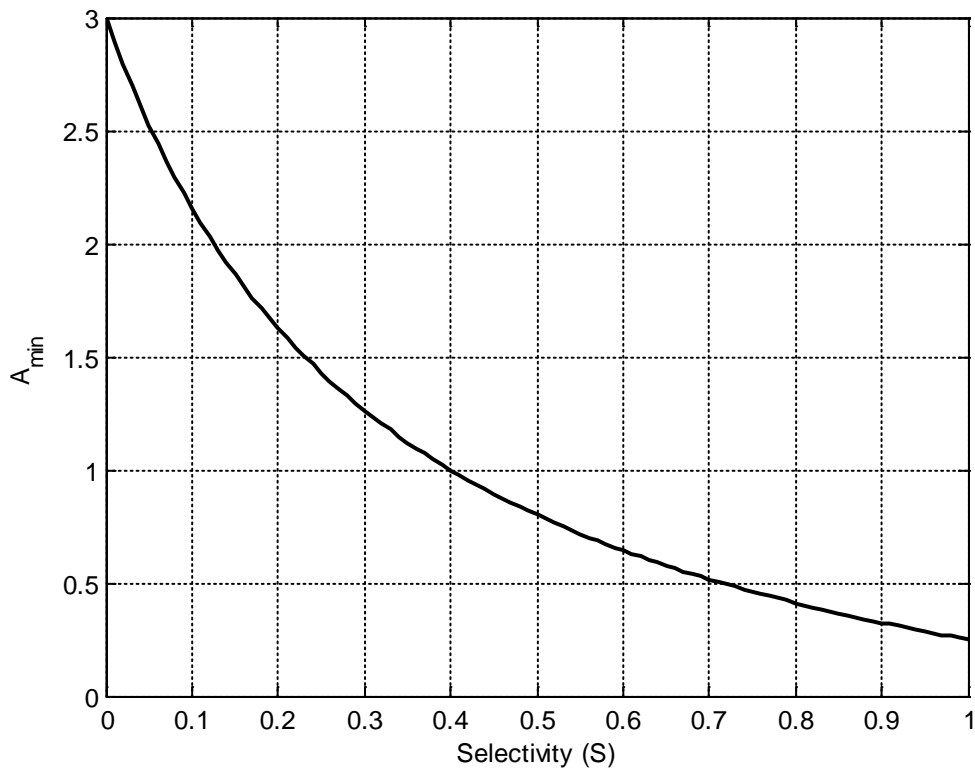

*Figure 8: Theoretical minimum of energy required for overall reaction in gas to liquid conversion.*

The gas conversion efficiency quantifies the amount of methane that will be incorporated into the liquid via one or both of the reactions 1 and 2 above. This depends on the experimental configuration and conditions, and can approach 50% efficiency.

In order to estimate the percentage increase in the mass of the liquid, we consider the $70{:}30$ composition described earlier.

In its *simplest form*, a single aromatic ring contains 6 carbons with unsaturated bonds (benzene). Thus, in the reaction where splitting of the carbon double bond leads to the addition of a methane molecule, the fragments $CH_3 \cdot$ and $H \cdot$ attach to separate $C$ atoms in the ring, occupying 2 sites. Consequently, a maximum of 3 methane molecules can theoretically react with a 6 carbon ring forming something that might look like 1-3-5-methylcyclohexane. In general, with multiply linked aromatic rings, the number of potential reactions for saturation by a methane fragments is $n/2$, where the number of carbon atoms is $n$. The mass of carbon is 12 times higher than hydrogen, thus the mass composition of the liquid is almost entirely due to the carbon component. Consequently, a liquid containing 70% aromatic composition has a maximum mass increase of 35% if only reactions of type 2 take place to fully saturate every molecule in the volume. Of course, 100% saturation of the ring is not likely, and we introduce the factor $\eta$ to represent the percentage of the aromatic ring that gets saturated on average.

Considering linear alkane chains of dodecane $C_{12}H_{26}$, the most likely positions for polymerization of the chain is at the ends of the molecule. Thus, for each molecule, we consider the existence of 2 sites for polymerization. With the same mass relationship as before, and 30% composition of saturated molecules in the liquid, this gives 5% increase in mass of the liquid by polymerization. Clearly this is a conservative estimate as it does not include extended polymerization process which can occur repeatedly on the same molecule.

The overall mass increase efficiency in the liquid can be expressed as:

$$\alpha(S) = S\eta\, 35\% + (1 - S)5\% \qquad (6)$$

The dependence of $\alpha$ on the reaction rate parameter $\eta$ is shown in Figure 9.

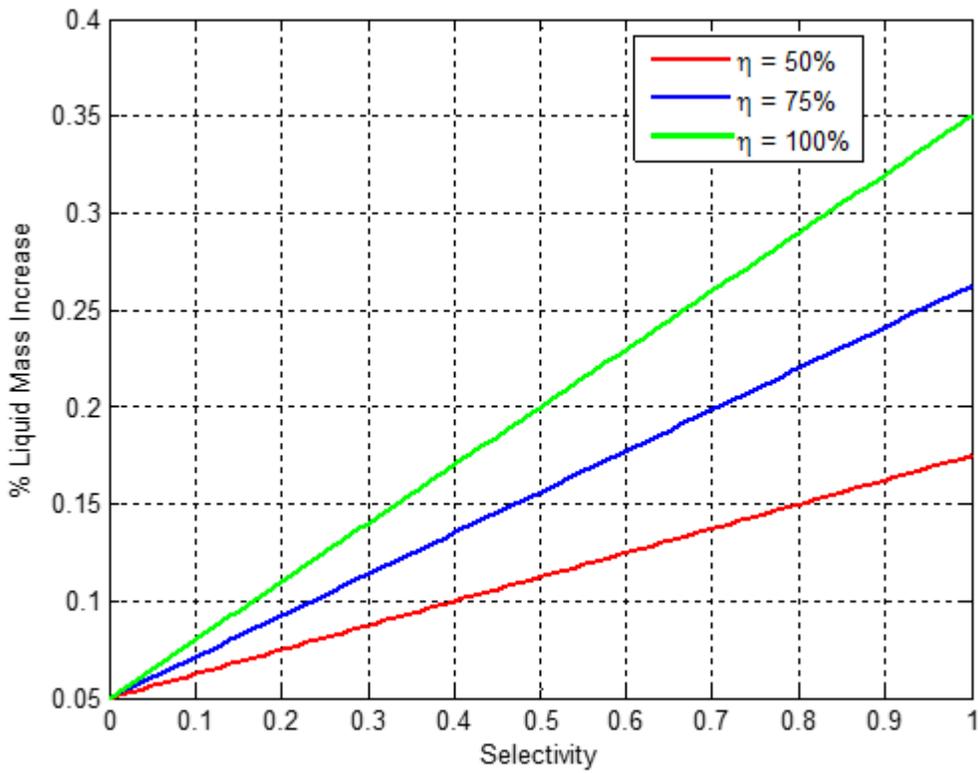

*Figure 9: Liquid mass increase as a function of reaction selectivity for different degrees of saturation on the aromatic ring.*

Finally, we can estimate the maximum temperature increase of the liquid under these conditions according to the relationship:

$$T_{max} = \frac{\Delta T}{C_p} = \frac{A_{min} - \Delta H}{C_p} \qquad (7)$$

Here, $C_p$ is the dimensionless specific heat of the liquid, taken as 5 for the liquid under consideration.

The maximum temperature increase as a function of $S$ is derived through interpolation of the data points at $S = 0$ and $S = 1$ which are the two limiting extremes of the reaction process.

When $S = 1$, the only mechanism is saturation of double bonds in the aromatic molecule, which according to Figure 8 required 0.25 eV for the process. Since the enthalpy change in the reaction is $-0.5$ eV/mol, the change in temperature is computed from equation (7):

$$T_{max} = \frac{0.25 - (-0.5)}{C_p} \alpha = 0.026 \text{ eV} \approx 304 \text{ K}$$

In the other limiting case, where $S = 0$, and polymerization is the only relevant process, the required energy from Figure 8 is 3 eV, for a process which has an enthalpy change of 0.6 eV/mol. Thus, the temperature increase in this case is given by equation (7) as:

$$T_{max} = \frac{3 - 0.6}{C_p} \alpha = 0.024 \approx 278 \text{ K}$$

Both of these values are computed for $\eta = 50\%$. The curves in Figure 10 show the change in liquid temperature over different values of $\eta$.

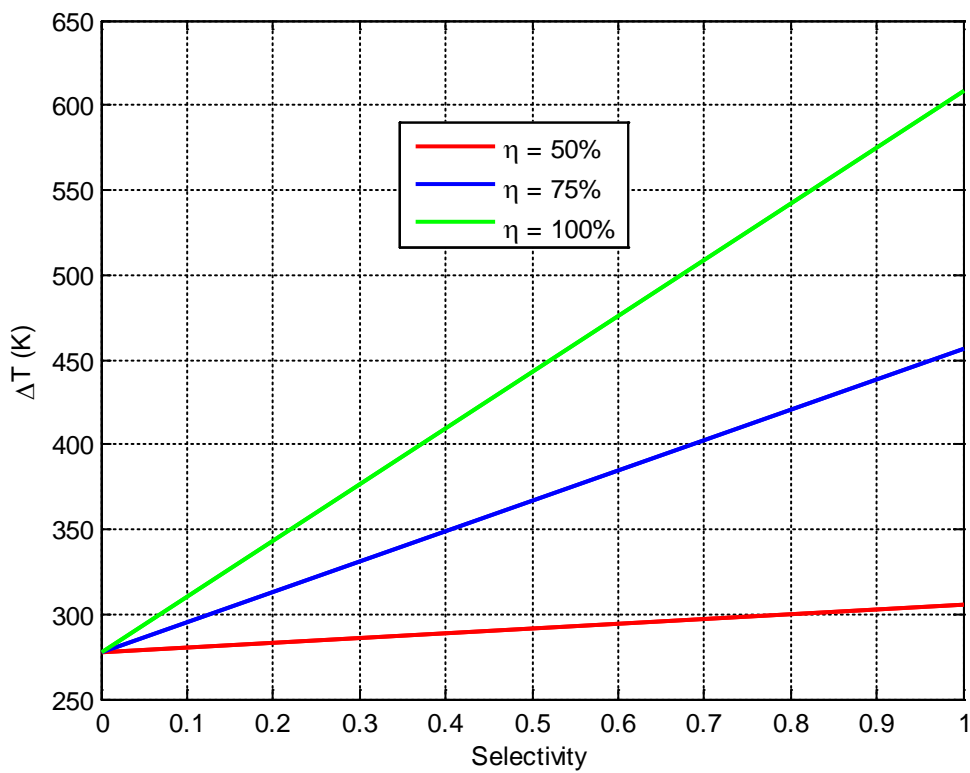

*Figure 10: Change in liquid temperature as a function of reaction selectivity for varying degrees of* saturation parameter ***η***.

## 4. Conclusions

A new approach of methane liquefaction by direct incorporation of excited methane into liquid hydrocarbons using nanosecond-pulsed dielectric barrier discharge and atmospheric pressure glow discharge has been tested. NMR and FTIR analysis of treated liquid samples as well as GC analysis of gas phase show fixation of methane in liquid fuel, decrease of methane concentration in gas phase, as well as structural changes in liquid hydrocarbons (saturation of double bonds and opening of hydrocarbon rings). Analytical model of this process is proposed. Although these results are potentially promising, further research is required, since many unanswered questions and gaps in understanding remain.

### Acknowledgments

This work was in part supported by Advanced Plasma Solutions (APS) and Drexel University.